# Structure motif centric learning framework for inorganic crystalline systems

Huta R. Banjade[a#], Sandro Hauri[b#], Shanshan Zhang[b], Francesco Ricci[c], Geoffroy Hautier[c], Slobodan Vucetic[b*], Qimin Yan[a*]

[a] Department of Physics, Temple University, Philadelphia, Pennsylvania 19122, USA

[b] Department of Computer and Information Science, Temple University, Philadelphia, Pennsylvania 19122, USA

[c] Institute of Condensed Matter and Nanoscience (IMCN), Université catholique de Louvain (UCLouvain), Chemin étoiles 8, bte L7.03.01, Louvain-la-Neuve 1348, Belgium

*Correspondence and requests for materials should be addressed to Q. Y. (qiminyan@temple.edu) or S. V. (vucetic@temple.edu).

[#] H.B. and S.H. contributes equally to this work.

**Abstract**

Incorporation of physical principles in a network-based machine learning (ML) architecture is a fundamental step toward the continued development of artificial intelligence for materials science and condensed matter physics. In this work, as inspired by the Pauling's rule, we propose that structure motifs (polyhedral formed by cations and surrounding anions) in inorganic crystals can serve as a central input to a machine learning framework for crystalline inorganic materials. Taking metal oxides as examples, we demonstrated that, an unsupervised learning algorithm *Motif2Vec* is able to convert the presence of structure motifs and their connections in a large set of crystalline compounds into unique vector representations. The connections among complex materials can be largely determined by the presence of different structure motifs and their clustering information are identified by our *Motif2Vec* algorithm. To demonstrate the novel use of structure motif information, we show that a motif-centric learning framework can be effectively created by combining motif information with the recently developed atom-based graph neural networks to form an atom-motif dual graph network (AMDNet). Taking advantage of node and edge information on both atomic and motif level, the AMDNet is more accurate than an atom graph network in predicting electronic structure related material properties of metal oxides such as band gaps. The work illustrates the route toward fundamental design of graph neural network learning architecture for complex material properties by incorporating beyond-atom physical principles.

**Introduction**

Machine learning (ML) methods in combination with massive material data offers a promising route to accelerate the discovery and rational design of functional solid-state compounds by utilizing a data-driven paradigm.[1] Supervised learning has been effective in materials property predictions, include phase stability,[2–4] crystal structure,[5,6] effective potential for molecule dynamics simulations,[7,8] and energy functionals for density functional theory based simulations.[9] With the recent progress in deep learning, ML has also been applied to inorganic crystal systems to learn from high-dimensional representations of crystal structures and identify their complex correlations with materials properties. For instance, band gaps of given classes of inorganic compounds have been predicted using deep learning[10,11] and ML has been applied on charge densities,[12,13] and Hamiltonian data[14] to predict electronic properties.

Recent development of graph convolutional network (GCN),[15–17] when combined with domain knowledges, offers a powerful tool to create an innovative representation of crystal structures for inorganic compounds. Within the GCN framework, any type of grid and atomic structure can be successfully modeled and analyzed. The flexible graph network structure endows these learning frameworks[15] a large room for improvement by considering more node/edge interactions in the crystal graphs.[18]

Whether ML can efficiently approximate the unknown nonlinear map between input and output relies on an effective representation of solid-state compound systems that captures structure–property relationships which form the basis of many design rules for functional materials. In inorganic crystalline materials with unit cells that satisfy the periodic boundary condition, bonding environments determined by local and global symmetry are essential components for the understanding of complex material properties.[19,20] As stated in the Pauling's first rule,[21] a coordinated polyhedron of anions is formed about each cation in a compound, effectively creating structure motifs that behave as fundamental building blocks and are highly correlated with material properties.

Structure motifs in crystalline compounds play an essential role in determining the material properties in various scientific and technological applications. For instance, the identification of $VO_4$ functional motif enabled the discovery of 12 vanadate photoanode materials via high-throughput computations and combinatorial synthesis.[22] In the field of complex oxide devices, $MnO_6$ octahedral motifs are correlated with small hole polarons that limit electrical conductivity.[23] In battery cathodes for energy storage, high ion mobility is explained by the local bonding environment of a multivalent ion.[24] $V^{4+}$ ion related motifs and the connections between these motifs are found to be important determining factors for the selective oxidation of hydrocarbons.[25–27] The presence of $MO_4$ tetrahedra ( M as Si or Al) can be used to identify the most promising synthetic candidates from the pool of hypothetical zeolites.[28] When designing novel battery materials, it is found that the changing coordination pattern of a migrating ion can be used as a descriptor of ion mobility.[29,30]

Governing the structure-property relationship, structure motifs or coordination environments can be viewed as effective structural descriptors for crystals. The efforts for identification of local coordination environments initially focused on structure types[31,32] or preferential coordination numbers[33] based on simple rules.[34,35] Very recently, owing to the development of data-driven approaches, systematic and robust approaches to automatically identify local environments have been developed,[36,37] which motivated the use of structure motif information for material design in a data-driven paradigm. For instance, structure motif information has been used to define crystal structure similarity for all the compounds in the Materials Project database.[38] A recent work comprehensively evaluated the validity and suggested the limited predictive power of the almost one-century-old Pauling rules.[39] Recent analysis and the dataset of local environment and connectivity[36] provide a novel set of material information that can serve as essential input for machine-learning techniques in materials science.

In this work, we propose to incorporate structure motif information in a machine learning framework. We show that the presence of structure motifs and their connections extracted from a material structure database can be utilized with unsupervised learning to define unique representations in a high-dimensional space. The dimension reduction process reveals strong clustering effects, representing the neighborhood properties of metal elements in the periodic table. By combining the motif information with graph convolutional neural networks, we develop a motif-centric deep learning architecture, called the atom-motif dual graph neural network (AMDNet), whose accuracy surpasses that of the state-of-the-art atom-based graph network MEGNet[17] for the prediction of electronic structures of inorganic crystalline materials.

**Motif2Vec: Structure Motif Vectorization**

In a recent work,[40] it is shown that an unsupervised learning algorithm called *Atom2Vec* can learn high dimensional vector representations of atoms that encode basic properties of atoms by utilizing an extensive database of chemical formulas. Clustering of atoms in the vector space classifies them into groups consistent with the periodic table. Furthermore, it is possible to use vector representations of atoms to calculate the similarities among materials and make property predictions. In this work, we will enhance the previous development by demonstrating that structure motifs encoded in crystal structures reveal useful information about structural properties and electronic structures of crystalline systems.

We focus on binary and ternary metal oxides that constitute a vast and diverse material space where crystal structures are well characterized by local environments through cation-oxygen coordination. The materials set includes 22,606 complex oxides in the Materials Project database.[41] We extract the structure motif information using the local environment identification method developed by Waroquiers *et al*.[36,42] as implemented in the pymatgen code,[43] following the definition of structure motifs or coordination environments by the International Union of Crystallography[44] and International Union of Pure and Applied Chemistry[45] as listed in Ref. 36.

We identify the connections between a motif and its neighboring motifs based on the number of oxygen atoms shared by those motifs. Three different types of connectivity may exist, from which we identify the connections as corner sharing (if only one atom is shared), edge sharing (if two atoms are shared) and face sharing (if three or more than three atoms are shared). The motif environment is defined by the neighboring motifs and the type of connection a motif has. By iterating through all the structures in the dataset, motif-environment pairs are identified and the motif environment matrix is generated. Details on the motif environment matrix are included in the SI.

Next, we propose the *Motif2Vec* algorithm that is able to take advantage of the above embedding process and convert each row of the motif environment matrix effectively into a high-dimensional vector that represents a unique structure motif. To create the vector representations for structure motifs, we treat motifs as the basic building blocks and study their presence and motif-wise environment in 22,606 oxide crystal structures extracted from the Materials Project database.

Figure 1 shows the high-level representation of the workflow used in the *Motif2Vec* machine. Material properties, such as orbital interactions within a crystal, are known to be related to bond lengths as well as bonding angles. We extract the following quantities to represent motif connections: (i) the distance between the cation center of a motif $M_1$ and its neighboring motif center ($M_2$); (ii) the $M_1$-O-$M_2$ bonding angles for those oxygen atoms shared by the two motifs. The extracted motif connection information will be an essential input for the learning process utilizing graph convolutional network as described below.

Our aim is to identify patterns and clustering information for these high-dimensional motif vectors that in turn influence the complex material properties of oxide compounds. By using various linear and nonlinear transformations, dimension reduction algorithms serve this purpose by creating a low-dimensional representation (called embedding) that best preserves the overall variance of the original dataset. To demonstrate the clustering of the motif vectors from *Motif2Vec*, we visualize the high dimensional data by using the t-SNE (t-Distributed Stochastic Neighbor Embedding),[46] a recently developed nonlinear dimensionality reduction technique that is well suited for visualization of high-dimensional datasets. Before the t-SNE, apply singular value decomposition (SVD)[47] to project the original high-dimensional representation of materials to 60-dimensions, corresponding to the largest 60 singular values. The detailed procedure for t-SNE is presented in SI.

Figure 2 shows the projected motif vector data in two dimensions (2D) obtained through the t-SNE process, where different motif types are represented by different colors. We observe that there exist distinct clusters based on the motif types. First of all, detailed analysis of those clusters shows that the chemical properties of the elements forming the motifs plays an important role in the formation of clusters. For instance, all the Lanthanide-based motifs formed different clusters on the basis of motif type (cluster 1 in Fig. 2 and cluster 9 in Fig. S3 in the SI). It is interesting to see that Yttrium-based motifs always stay close to Lanthanide-based ones, as the chemical properties of Yttrium are known to be similar to Lanthanides. In addition, motifs associated with Zn and Mg always cluster together, which is consistent with the fact that Zn is chemically similar to Mg because both of them exhibit only one normal oxidation state (+2) and their ions ($Zn^{2+}$, $Mg^{2+}$) are similar in size.

As shown in Figure 2, cluster 1 contains cubic motifs associated with Lanthanides, while the cuboctahedral motifs associated primarily with main group elements appear in cluster 2. The clustering of motifs determined by elements as described above is in accordance with the grouping pattern in the periodic table, although no information about the periodic table was used in the vectorization process. It is interesting to see that octahedral motifs associated mostly with the transition metal elements occur together in cluster 3, while the tetrahedral and square planer motifs associated with transition metal elements are located but well separated in cluster 4. This motif cluster separation demonstrates the power of *Motif2Vec* to capture local structural information as well as elemental information. Additional motif clusters in Figure 2 are presented in the SI (Figure S3). These findings achieved by unsupervised learning strongly support our intuition that structure motifs can serve as essential fingerprints for crystalline compounds that carry both elemental and structural information.

**Incorporation of Motif Information in Graph Neural Networks**

As above-atomic-level building blocks of crystals, structure motifs and motif-wise interactions within a crystal strongly influence the material properties. Structure motif information can be used as an essential input to a graph neural network that predicts physical properties of materials. Following the standard notation used in the graph neural network (GNN) framework,[48] we represent an attributed graph as $G = (V, E)$, where $V = \{v_i\}_{i=1,2,...,N^v}$ is a set of nodes of cardinality $N^v$, $v_i$ is the node attribute vector of the $i^{th}$ node. $E = \{(e_k, r_k, s_k)\}_{k=1,2,...,N^e}$ is a set of edges of carnality $N^e$, where $e_k$ is the attribute vector for edge $k$ between nodes $s_k$ and $r_k$. Several graph neural networks have been proposed[16–18] that formulate the task of predicting chemical properties of materials as learning a mapping $f(G:W) \rightarrow y$, where $W$ is a set of learnable parameters and y is a target property.

Most of the graph networks applied to crystalline materials[16–18] are based on graphs on the atomic level $G_0^{atom}$ as input for the network. Such atomic graphs contain information about atoms (such as atomic number, electronegativity and many others) and bonds. For instance, in the $G_0^{atom}$ of atomic graph network MEGNet, $v_i$ is a vector representing the $i^{th}$ atom in a unit cell and is represented by the atomic number of the element. $e_{ij}$ is a vector representing a bond between atom $i$ and atom $j$.

In this work, to enable a learning architecture that synthesize both atom-level and motif-level graph representation of materials, we propose that atom-motif dual graph networks can be constructed to enhance the learning process and improve the prediction accuracy for electronic structure properties of metal oxides. We follow the procedure introduced in existing atomic graph networks[17,49] to represent the edges, where two atoms are connected if they are no more than 5 Å apart. We propose to represent the metal oxides as motif graphs $G_0^{motif}$, where each motif in a crystal is represented by a node $\{v_i\}_{G_0^{motif}}$ and each connection between two motifs is represented by an edge $\{e_{ij}\}_{G_0^{motif}}$ as shown in Figure 3. Motif graphs represent the same materials with higher granularity than atom graphs, but more comprehensive information can be encoded in each motif node such as local distortions and site symmetries. The motif graph uses the same edge representation as in the atom graph, and the motif-motif edge distances are measured from the center atom of one motif to that of a neighboring motif.

In the motif graph, a combination of atom-level and motif-level information is encoded in each node. We adopt the atom-level node representations by combining two existing approaches to form a 103-dimensional vector that uses the information of atoms within the motif. The first 86 dimensions represents the fractional encoding of the atoms proposed by Meredig *et al.*,[50] and the next 17 dimension is for physical properties proposed by Ward *et al.*[51]. On the other hand, we define the motif fingerprints by order parameters (of dimension 61) which describes the numerical measure of the local environment around an atom relative to a target standard motif.[37,52] These 61 dimension vectors are then concatenated with 103-dimensional atom based feature to form final 164 dimensional vector. Detailed descriptions about various types of order parameters and methods to compute such parameters are presented in the work by Zimmermann *et al.*[53] All the structural information used to construct the motif graph, including extended connectivity, angle, distance and order parameters for each motif, are computed by using the python package *robocrystallography*[54] combined with the *pymatgen* code. By combining atomic-level and motif-level information, we utilize a 164-dimensional vector to represent each motif in the graph.

**AMDNet: Atom-Motif Dual Graph Neural Network**

A high-level illustration of our proposed atom-motif dual graph neural network (AMDNet) architecture is shown in Figure 4. To incorporate the motif information acquired above into the graph network learning framework, the central concept in the proposed architecture is to generate both motif graphs and atom graphs representing the same compounds, with different cardinality of edges and nodes, and combine the representation information before making predictions.

For each material, we generate an atom graph and a motif graph (Figure 4). We adopt the convolution structure of the MEGNet proposed by Chen *et al*.[17] when constructing the atom-level graph network. The choice of graph network structure is only for a benchmark purpose, and many other types of crystal graph convolution networks could be used to take advantage of the motif-level graph information.[16,18] As a preliminary test, we use the same architecture as that for the atom graphs in MEGNet to generate $G_0^{motif}$ by utilizing the 164-dimensional atom-motif-mixed vector input for the nodes in the network. Edges in $G_0^{motif}$ are defined as the distances between the center atoms of any two motifs. Note that MEGNet can be interpreted as a neural network that encodes the whole crystal graph input to a low dimensional vector of dimension 16, upon which a final single-value prediction is made. Taking advantage of this fixed-dimension representation of any MEGNet graph convolution network, we can effectively combine the information from motif and atom dual graphs by concatenating the two low-dimensional representations generated from motif graph and atom graph, respectively. This concatenated vector is then feed to a small feed-forward neural network for single-value predictions; more details are presented in the Methods section

We use 22606 binary and ternary metal oxides from the Materials Project database to evaluate the efficiency of our proposed model and focus on the prediction of band gaps which is one of the complex electronic structure problems. Metal oxides are a class of solid-state compounds that are challenging for both *ab initio* quantum simulations and machine learning in general, which is verified by our experiments on different datasets as presented in the SI. For the comparison purpose, we create a motif graph network model, MNet, which use motif graphs ($G_0^{motif}$) as the only input to the network. Table 1 shows a comparison between MEGNet, MNet, and our proposed AMDNet on the prediction accuracy of band gaps, formation energy, and the metal (compounds with band gaps less than 0.2 eV in the MP database) vs. non-metal (compounds with band gaps 0.2 eV or greater) classification for all the metal oxides in our dataset.

The results show that, given the same training and test data, AMDNet shows its superiority in the band gap prediction task in compared to the state-of-the-art baseline model. The motif graph representation (MNet) performs worse than MEGNet, which is expected because it uses a much smaller graph representation. The combination of atom and motif graph (AMDNet) outperforms MEGNet on the band gap prediction task, which illustrates that the motif representations enhance the effective learning of material properties. Figure 4(b) shows the comparison of the predicted band gaps on the test dataset with the actual band gaps. Band gaps of a large portion of the compounds are clustered close to the diagonal, indicating a good performance of our model on the band gap prediction task. In addition, our model shows superior performance in the metal vs. non-metal classification task. As shown in Table 1, the classification accuracy is 82.1% for AMDNet while for MEGNet it is only 75.3%. On the other hand, the formation energy prediction shows almost identical performance with MEGNet, indicating atom graph alone is sufficient for the formation energy prediction task, which is considered as a simpler task in compared to the band gap prediction task. The comparison between predicted (by AMDNet) and actual formation energies is shown in Figure 4(c), and the comparison of prediction accuracy given by various models is shown in Table 1.

**Summary and discussion**

We demonstrate in this work how structure motifs in crystal structures can be combined with both unsupervised and supervised machine learning techniques to enhance the effective representation of solid-state material systems. As a step forward from *Atom2Vec* to *Motif2Vec*, motif vectors learned from motif environments in 22606 metal oxides using unsupervised learning effectively capture the motif similarities and their clustering properties. To enhance the learning of solid-state crystalline systems for complex electronic structures, structure motif and connection information are incorporated as essential input in an atom-motif dual graph neural network model (AMDNet), which outperforms the state-of-the-art atom graph neural network model for the prediction of electronic band gaps and metal vs. non-metal classification task. In addition, AMDNet model is able to predict formation energy in close agreement with the existing state-of-the-art atom graph-based models.

AMDNet is a general learning framework for solid-state atomistic systems that can be used to predict other materials properties, such as mechanical and excited state properties, and applied to other motif-based systems including two-dimensional materials and metal-organic frameworks.

Several directions related to the motif-centric learning methods here are worthy to explore in the future. Although we perform the test on perfect crystalline systems, through the addition of extra types of local motif information, the motif-enhanced graph network framework can be expanded for the learning and prediction of surface and defective material systems. Besides the use of a dual graph network architecture, motif information and the physical principles behind it can be incorporated into a learning framework in other manners, such as through a motif-enhanced convolutional process in an atom-based graph convolutional network or other novel algorithms that are actively developing in the graph theory including graph attention.

**Methods**

*Training process for atom motif dual graph neural network:* In the AMDNet with $L$ layers, the module generates a sequence of atomic graph representation $\{G_1^{atom}, G_2^{atom}, ...., G_L^{atom}\}$ and motif graph representations $\{G_1^{motif}, G_2^{motif}, ...., G_L^{motif}\}$, where each graph has the same number of nodes and edges as in the input graphs $G_0^{atom}$ and $G_0^{motif}$, respectively. Through a graph convolutional process called AtomNet Block for atom graphs and MotifNet Block for motif graphs information of each edge and its respective connecting nodes are passed through a dense neural network with a nonlinear activation function (we use the shifted softplus function), which creates a new edge representation. To generate the new node representation, the node information together with the information of the new incident edges is passed through a separate dense neural network with the same nonlinear activation function.

Each graph convolutional block has a hidden dimension of 64 for both node and edge convolution. In our work, we use three graph convolutional blocks to apply the graph convolution, which creates an output graph representation. The graph representation is transformed into vector form by averaging over all nodes and edges respectively, which is denoted as set2set(E) and set2set(V) in Figure 4(a). These set2set vectors are concatenated before going through two densely connected layers as shown in Figure 4(a). This results in a low-dimensional vector representation of the original atom and motif graph representation of the crystal. These representations are concatenated again and passed through two densely connected layers to make a single real-valued prediction.

For the training and test process, we choose a 60-20-20 train-validation-test splits. We initialize the hyperparameters based on the best values from MEGNet to train our neural network. All deep models are trained with ADAM optimizer[55] with initial learning rate α = 0.001. Training formation energy prediction was slower to converge to the best solutions than for the band gap prediction, therefore we adjusted some parameters to adapt to each prediction task. We stop training when the validation error doesn't improve for 20 and 100 epochs to train band gap prediction and formation energy prediction, respectively. We save the model with the lowest observed validation error and use it to evaluate the models on the test data. We use 64 compounds per minibatch for band gap prediction and 32 compounds per minibatch for formation energy prediction.

**Acknowledgments**

H. Banjade and Q. Yan acknowledge support from the U.S. Department of Energy under Award number DE-SC0020310. F.R. acknowledges financial support from F.R.S - FNRS. It benefitted from the supercomputing resources of the National Energy Research Scientific Computing Center (NERSC), a U.S. Department of Energy Office of Science User Facility operated under Contract No. DE-AC02-05CH11231.

Table 1. Performance comparison between various graph architectures for the learning and prediction of electronic band gaps, formation energy per atom, and metal vs. non-metal classification accuracy for the metal oxides (trained on 18,091 compounds and tested on 4,515 compounds).

| Model | Band gap MAE (eV) | Formation energy MAE (eV/atom) | Metal vs nonmetal classification accuracy |
|---|---|---|---|
| MEGNet (atom graph) | 0.542 | 0.0469 | 75.3% |
| MNet (motif graph) | 0.639 | 0.1214 | 74.7% |
| AMDNet (motif-atom dual graph) | 0.443 | 0.0470 | 82.1% |

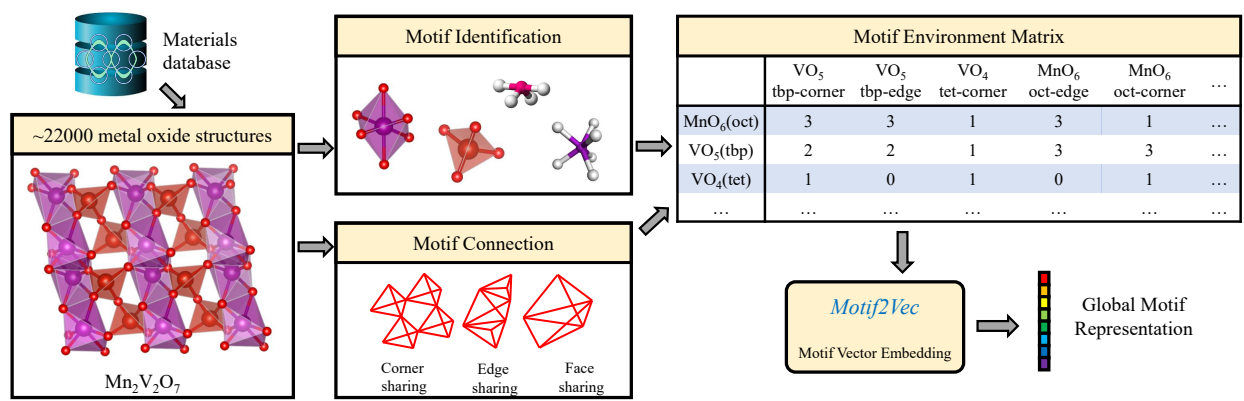

**Figure 1** Extraction of structure motif information in inorganic crystalline compounds (metal oxides) and the generation of global motif representations by *Motif2Vec* machine using the motif environment matrix.

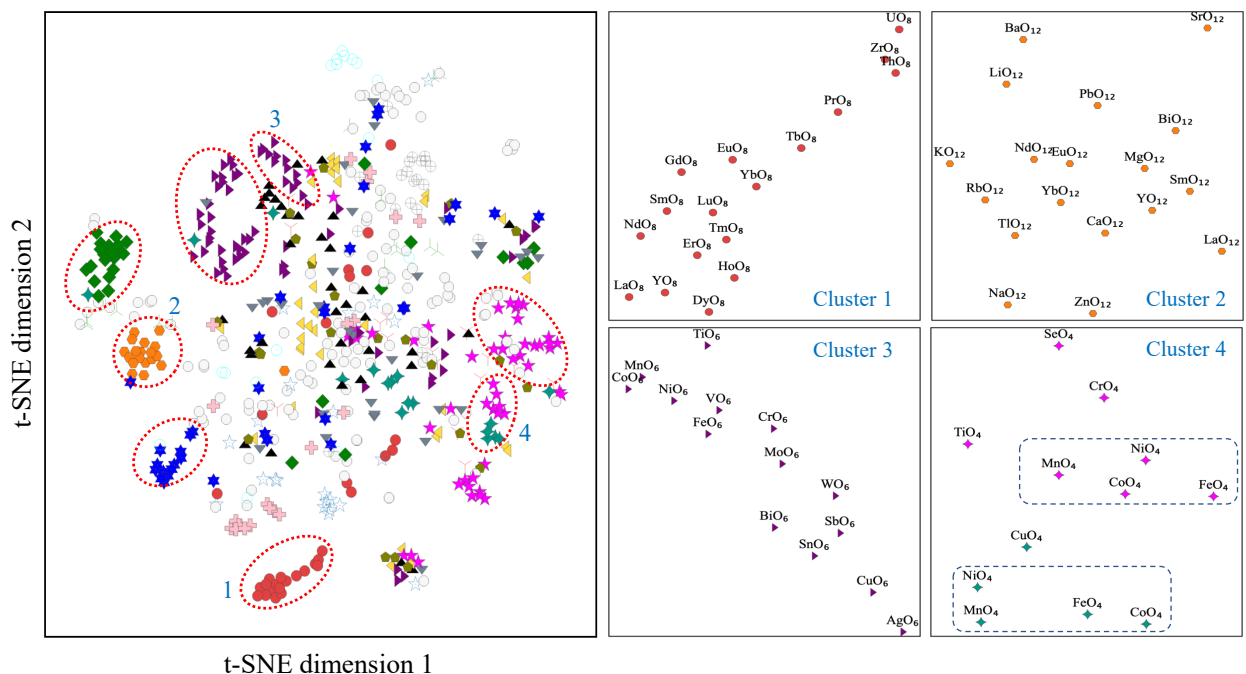

**Figure 2** t-SNE projection of motif vectors constructed by using the motif environment matrix. The motif clusters 1 to 4 are associated with various motif types including (1) cube, (2) cuboctahedron, (3) octahedron, and (4) a mixture of tetrahedron (in magenta) and square plane (in remnant).

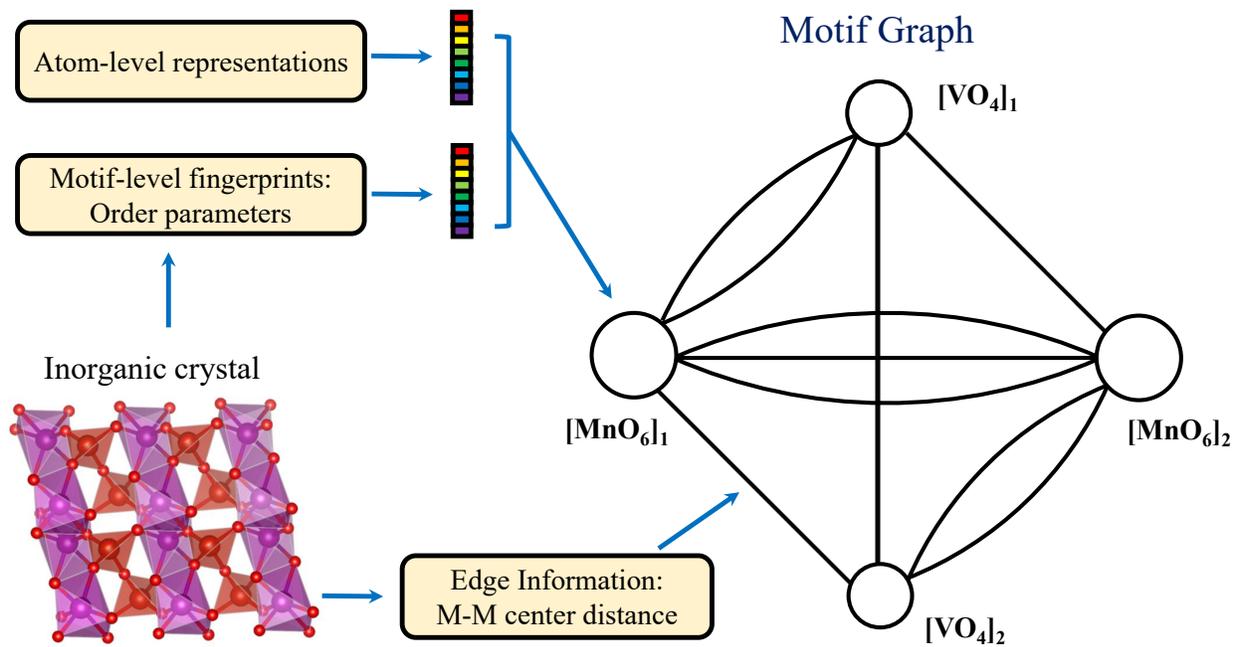

**Figure 3** Construction of a motif graph based on both atom-level and motif-level information encoded in an inorganic crystal.

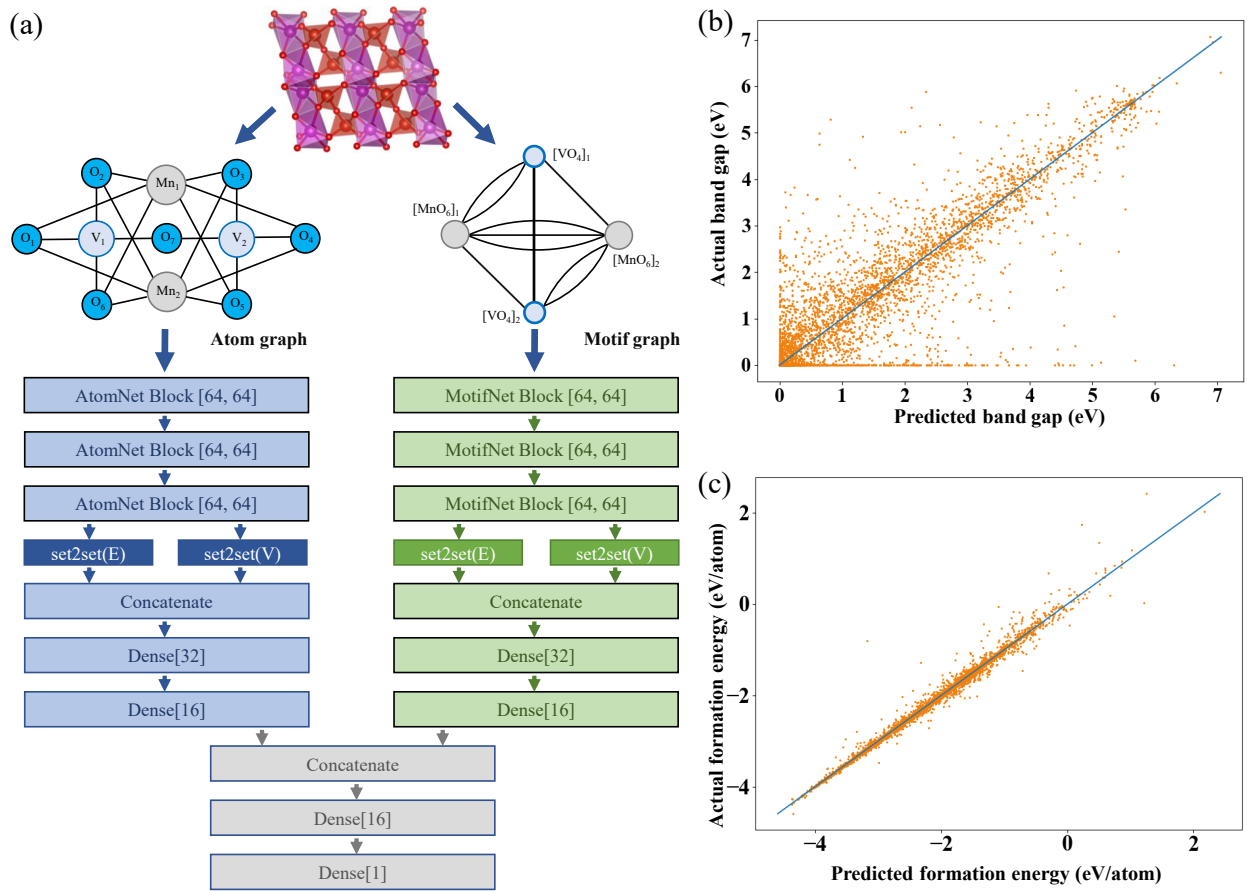

**Figure 4** (a) Demonstration of the learning architecture of the proposed atom-motif dual graph network (AMDNet) for the effective learning of electronic structures and other material properties of inorganic crystalline materials. (b) Comparison of predicted and actual band gaps (from DFT calculations) and (c) comparison of predicted and actual formation energies (from DFT calculations) in the test dataset with 4515 compounds.

# Structure motif centric learning framework for inorganic crystalline systems

Huta R. Banjade[a#], Sandro Hauri[b#], Shanshan Zhang[b], Francesco Racci[c], Geoffroy Hautier[c], Slobodan Vucetic[b]*, Qimin Yan[a]*

[a] Department of Physics, Temple University, Philadelphia, Pennsylvania 19122, USA

[b] Department of Computer and Information Science, Temple University, Philadelphia, Pennsylvania 19122, USA

[c] Institute of Condensed Matter and Nanoscience (IMCN), Université catholique de Louvain (UCLouvain), Chemin étoiles 8, bte L7.03.01, Louvain-la-Neuve 1348, Belgium

* Correspondence and requests for materials should be addressed to Q. Y. (qiminyan@temple.edu) or S. V. (vucetic@temple.edu).

[#] H.B. and S.H. contributes equally to this work.

**Identification of structure motif types**

Among different neighbor finding modules available in pymatgen[1] code, we use the SimplestChemenvStrategy module (with the control parameters "distance-cutoff" = 1.4, "angle-cutoff" = 0.3, and "continuous-symmetry-measure-cutoff" = 10), which uses a fixed distance and the angle parameters to identify the nearest neighbors of any site in a crystal. To be consistent with the definition of a cation-anion structure motif, in addition to the distance and angle cutoff, we consider only the cation-anion bonding and ignore the bonding between the same elements (additional condition = 3 in SimplestChemenvStrategy in pymatgen). Once the neighboring atoms are identified, continuous symmetry measure (CSM) is used to measure the similarity between a given local environment and the perfect environment.[2-4] Through this process, a local motif can be uniquely defined and assigned a coordination environment type that corresponds to the lowest value of continuous symmetry measure.

Note that it is a very difficult task to assign exact motif types for all sites in a large set of compounds using a uniform set of control parameters. One would always expect some deviations from the ground truth by looking at the structures of some compounds through human eyes. We observe that the approach presented by Waroquiers D. *et al.*,[5] returns a higher percentage of correct motif types while doing large scale analysis. More details regarding the neighbor finding approach and the available strategies can be found in Ref. 5.

**Calculations of order parameters and motif connectivity**

We identify the nearest neighbors of all sites in a crystal and generate the bonded structure graph by using CrystalNN module available in the pymatgen code. This bonded structure graph is then analyzed by using the robocrystallography[6] module available in the pymatgen code, and the information regarding the site geometry, bonding distance, and the motif connectivity type (corner, edge, or face sharing) within the unit cell and its neighboring cells are extracted. The motif-level site fingerprint is determined by order parameter which is computed with the module robocrystallography as a 61-dimensional vector.[7] It contains information about the local coordination environment and its corresponding weight ("wt"). The "wt CNx" in such a vector implies how much the environment resembles a certain coordination number (regardless of geometrical arrangement). Other parameters are obtained by multiplying the "weight" for that coordination number with the order parameter value. For example, "wt $CN_2$" provides a 2-fold likelihood while the "L-shaped CN$_2$" provides similarity to an L-shaped coordination geometry. The detailed analysis of such local environments and the complete list are presented in Ref. 6 and Ref. 7 respectively. The abbreviations we used for such local environments are presented in the last page of this supporting document.

**Details on the motif environment matrix**

By using the pymatgen code and the method described in Ref. 5., all the motif types associated with cations in a complex oxide are determined, in addition, the connectivity type of each motif site is assigned, which is calculated by using the method describe in previous section. By iterating over all the compounds in our dataset (22606 complex oxides), all the motif-environment pairs are identified and a motif environment matrix (*M*) is constructed. Each entry $M_{ij}$ in M represents the frequency of connection of the i$^{th}$ motif-type with its j$^{th}$ motif environment.

|  | Environment | | | | | |
|---|---|---|---|---|---|---|
| Motif & its type | VO$_5$(tbp-corner) | VO$_5$(tbp-edge) | VO$_4$(tet-corner) | MnO$_6$(oct-edge) | MnO$_6$(oct-corner) | TeO$_6$(oct-edge) ...... |
| MnO$_6$(oct) | 3 | 3 | 1 | 3 | 1 | 0 |
| VO$_5$(tbp) | 2 | 2 | 1 | 3 | 3 | 0 |
| VO$_4$(tet) | 1 | 0 | 1 | 0 | 1 | 0 |
| SiO$_4$(tet) | 0 | 0 | 0 | 0 | 0 | 0 |
| TeO$_6$(oct) | 0 | 0 | 0 | 0 | 0 | 1 |
| FeO$_6$(oct) | 0 | 0 | 0 | 0 | 0 | 0 |
| CoO$_6$(oct) | 0 | 0 | 0 | 0 | 0 | 0 |
| ⋮ | ⋮ | ⋮ | ⋮ | ⋮ | ⋮ | ⋮ |

**Figure S1** An example of the motif-environment matrix (M), a sample dataset of 5 compounds (Mn$_2$V$_2$O$_7$ (mp-19142, mp-1221818, mp-572632), Si$_2$TeO$_6$ (mvc-4099), Fe$_5$Co$_3$O$_{16}$ (mp-771671)) is used to generate this matrix. Each entry in M represents the frequency of the motif-environment pairs in the sample dataset. Abbreviations used: Octahedral (oct), Trigonal bipyramid (tbp), Tetrahedral (tet).

A schematic of *M* for a small set of compounds is shown in Figure S1. Each column in *M* gives the count of different environments with a single motif type and each row gives the count for different motif types with a single environment. Two motif types will behave similarly if their corresponding row vectors are close to each other in the high-dimensional vector space. As each motif type is related to only a small portion of all environments, M is extremely sparse and contains a very high dimension (our motif matrix contains 4373 rows and 10264 columns). The motif environment matrix thus constructed is analyzed and preprocessed before performing any further analysis. First, the entry $M_{ij}$ which contains the motif type and the environment with the same element as the center of the motif (For example: $MnO_6$(oct) as the motif type and $MnO_6$(oct-corner) as an environment) is assigned a zero value as it introduces the unevenness in the dataset. In the next step, to choose the dominant motif types in the material set, we consider the sum of entries of any motif type and it's all environments greater than or equal to 20. With this criterion, we obtain 610 most dominant motif types, which are used for the unsupervised learning task. The sum of counts over any row gives the total frequency of a motif type with all other environments available, which differs greatly among all motif types. Such an unevenness in frequency distribution is treated by a normalization process, $\mathcal{M}_{i,j} = M_{i,j}/(\sum_j M_{i,j}^p)^{\frac{1}{p}}$ on row vectors. The relative importance between dense and sparse environments can be tuned by using the hyperparameter p. Among different choices for p, we choose p = 2 as it gives the natural distance measure between different vectors.

**t-SNE method**

In the original work by L. van der Maaten *et al.*,[8] it is suggested that dimensionality reduction, such as principal component analysis (PCA)[9] or singular value decomposition (SVD)[10] should be performed before t-SNE on high-dimensional datasets. This helps to speed up the computation of the pairwise distance between data points and suppress some noise without severely distorting the interpoint distance. In this work, we performed the SVD on the motif environment matrix and obtained top 60 dimensions with the help of scree plot (as shown in Figure S2) for further analysis.

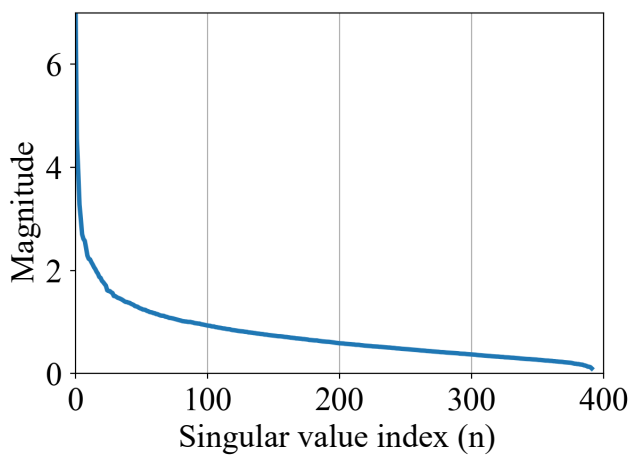

**Figure S2** Distribution of singular values obtained from the motif-environment matrix (with rows having sum >= 20) in descending order. The flat tail of non-vanishing values follows a sharp

decrease for the first ~50 dimensions. It indicates that higher dimensions of singular values could also describe the meaningful aspects of motifs. In this work, we choose motif vectors with singular values up to dimension 60 for further analysis.

The most important feature, while performing t-SNE is the perplexity; which is a cost function parameter to control the range of neighborhoods used in computing the probability distribution. As an important parameter on the resulting visualizations, perplexity is an estimate about the number of nearest neighbors considered when matching the high dimensional and low dimensional distribution for each point. By comparing the resulting visualizations with different values of perplexity we identified that a value of 30 is the best choice for our dataset. Other parameters that control the quality of the resulting embedding are the learning rate, the maximum number of iterations, and the optimization method used. We choose the learning rate as 200, the maximum number of iterations for the optimization is set to 15000, and the default method "exact" is used for the gradient calculations. All of the dimensional reduction and the visualization task in this work are done by using Scikit-learn,[11] a machine learning library for python programming.

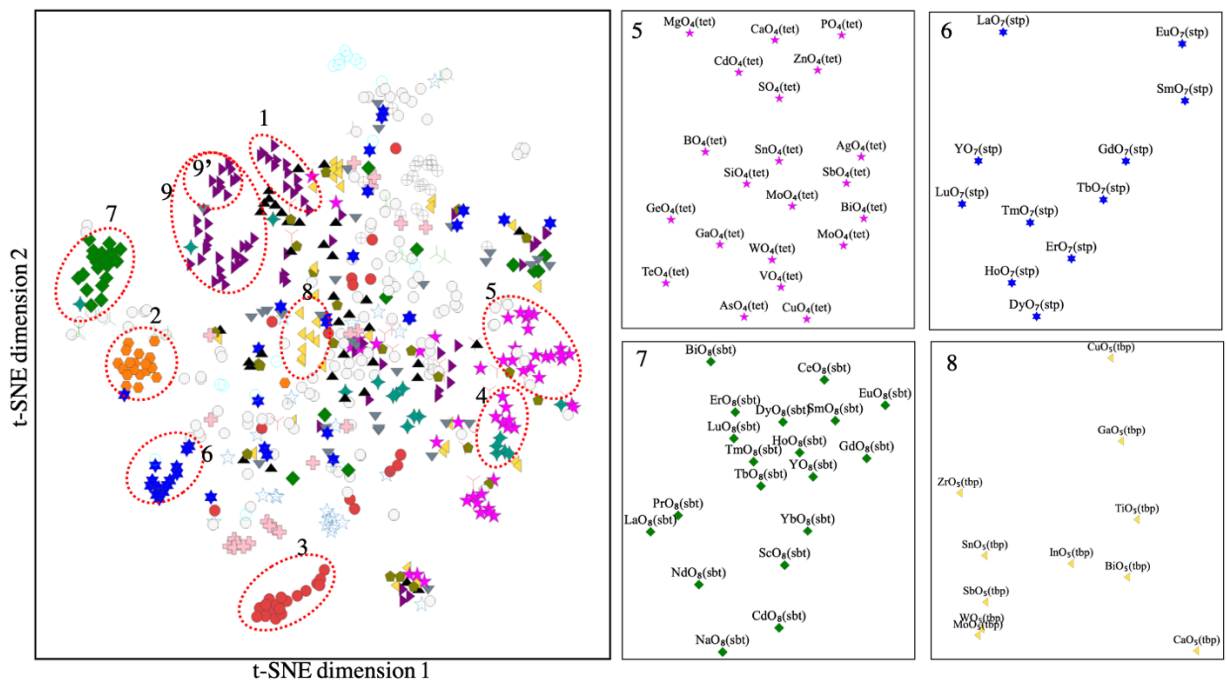

**Figure S3** t-SNE plot for the motif environment matrix, additional clusters 5, 6, 7, and 8 are presented. Cluster 9 and 9' are associated mainly with in octahedral motifs based on lanthanides and transition metal elements.

**Evaluation of band gap prediction for metallic and non-metallic oxides**

Chen *et al.*'s work focused on non-metallic (i.e., compounds with non-zero band gap) materials in the Materials Project database for the band gap prediction task, while our focus is binary and ternary metal oxides that are either metallic (compounds with band gaps less than 0.2 eV) or non-metallic (compounds with band gaps greater or equal to 0.2 eV). This requires utilizing a significantly different material dataset, and the direct comparison between our general experiments - especially on band gap prediction - and the published results from Chen *et al.*'s work is not possible. To ease the comparison, we split our material dataset with known band gaps into three parts: non-metallic non-oxides (26455), non-metallic oxides (15434) and metallic oxides (7172). While we focus on the prediction of oxides, the non-metallic non-oxides will help to compare our results with Chen *et al.*'s work. The non-metallic oxides are the overlap data between ours and Chen *et al.*'s. We use stratified sampling by generating the train-validation-test splits on each of those groups, which allows us to compare the performance on different subsets of the data and allows us to further investigate the performance of the proposed predictors. Metals have a band gap close to or equal to zero, while non-metals can vary a lot in the values of band gaps.

**Table S1** Comparison of the performance of MEGNet model on different subsets of the data. Predicting the band gap only non-metallic non-oxides leads to a smaller MAE than trying to predict the band gap of metal and non-metal oxides.

| Model | Band gap MAE (eV) | | |
|---|---|---|---|
| | Oxides and non-oxides (41889 non-metallic compounds) | Oxides only (15434 non-metallic compounds) | Oxides only (22606 metallic & non-metallic compounds) |
| MEGNet (atom graph) | 0.307 | 0.520 | 0.542 |

We recreated the experiment by Chen *et al.* as shown in Table S1 and found an even better result for band gap prediction on non-metallic materials than what is published in their work. This is mainly because they pre-trained the network for formation energy prediction and only retrained the last three densely connected layers to do band gap prediction, while we trained all layers from scratch to specialize on the target prediction. However, when we train the MEGNet to predict band gaps of (both metallic and non-metallic) oxides, the MAE is much higher. It indicates that band gap prediction for oxides is a more difficult prediction task. Even when we only consider non-metallic oxides, the error is much larger, showing the challenge of predicting the band gap of oxides. This observation is as expected, as many of these oxides belong to the so-called complex oxide class with transition metal elements that host multiple charge states and/or controlled by strong correlation interactions. Furthermore, electronic structures of these complex oxides are very sensitive to local lattice distortions and the connections of structure motifs. The band gaps of these oxides are thus correlated with complicated electronic structure information that is challenging for machine learning predictions through a structure-property mapping.

**Table S2** Test MAE on different subsets of the test data. Additional to the result in Table 1, we evaluate separate test errors for on non-metal (non-zero band gap) and metal (zero band gap) oxides. Training was done on all the oxides (22606 compounds).

| Model | Band gap MAE (eV) all oxides | Band gap MAE (eV) non-metallic oxides | Band gap MAE (eV) metallic oxides |
|---|---|---|---|
| MEGNet (atom graph) | 0.542 | 0.524 | 0.579 |
| AMDNet (atom-motif dual graph) | 0.443 | 0.442 | 0.443 |

Table S2 shows that the MAE for MEGNet on metallic oxides is roughly 10% bigger than on non-metallic oxides, which suggests that it struggles to properly differentiate between the two categories. As a result, a significant amount of metallic oxides are predicted as nonmetallic (see also Figure 4 in the main text). Our proposed AMDNet has roughly the same MAE for metallic and non-metallic oxides. This means that it learns to distinguish between those two categories, while giving an equally accurate estimate on the non-metallic oxides.